\documentclass[reqno]{amsart}
\usepackage{graphicx,amsmath,color}
\usepackage[toc,page]{appendix}
\usepackage{url}

\newcommand{\eps}{\varepsilon}

\newcommand{\pt}[1]{\partial_t #1}

\newcount\n  \newdimen\w

\def\Repeat#1#2{\n=#1\relax\loop\ifnum       
  \n>0\relax #2\advance\n by-1\repeat}

\long\def\OMIT#1{\relax }  

\def\re#1{(\ref{#1})}   
  
\def\eqn#1#2{ \begin{align} \label{#1}         #2 \end{align}}

\def\nnl#1{ \tag*{} \\ \label{#1}        }  

\def\delim#1#2#3{\csname\ifcase#1 relax\or   
   big\or Big\or bigg\or Bigg\fi\endcsname   
  {\ifcase#2\or\Delim#3\or\deliM#3\fi}}      
\def\Delim#1{\ifcase#1\relax\or(\or[\or\{\or<\or\langle\or|\or\|\or---{ }\fi}
\def\deliM#1{\ifcase#1\relax\or)\or]\or\}\or>\or\rangle\or|\or\|\or{ }---\fi}

\def\largerfrac#1#2#3{      
  \whichtypesize\n=\currenttypesize\advance\n by #1 \mathchoice
  {\setbox0\hbox{$\displaystyle-$} \w=.5\ht0\advance\w by-.5\dp0\setbox0
    \hbox{\typesize\n $\displaystyle-$} \advance\w by -.5\ht0\advance\w
    by .5\dp0\raise\w \hbox{\typesize\n$\displaystyle{\frac{#2}{#3}}$}}
  {\setbox0\hbox{$-$} \w=.5\ht0 \advance\w by -.5\dp0 \setbox0\hbox
    {\typesize\n $-$} \advance\w by-.5\ht0\advance\w by
    .5\dp0\raise\w\hbox{\typesize\n$\frac{#2}{#3}$}}
  {\setbox0\hbox{$\scriptstyle-$} \w=.5\ht0 \advance\w by-.5\dp0\setbox0
    \hbox{\typesize\n $\scriptstyle-$} \advance\w by -.5\ht0 \advance\w
    by .5\dp0 \raise\w\hbox{\typesize\n$\scriptstyle{\frac{#2}{#3}}$}}
  {\setbox0\hbox{$\scriptscriptstyle-$} \w=.5\ht0
    \advance\w by -.5\dp0 \setbox0\hbox{\typesize\n
    $\scriptscriptstyle-$} \advance\w by -.5\ht0 \advance\w by .5\dp0
    \raise\w\hbox{\typesize\n$\scriptscriptstyle{\frac{#2}{#3}}$}}  }

\def\d{{\rm d}}       

\def\Laplace{\bigtriangleup}  
\def\bg{{\bf g}} 
\def\bx{{\bf x}} 
\def\bv{{\bf v}} 
\def\bj{{\bf j}} 
\def\bJ{{\bf J}} 
\def\bE{{\bf E}} 
\def\bb{{\bf b}} 
\def\bW{{\bf W}} 
\def\bV{{\bf V}} 
\def\bId{{\bf I}} 
\def\bP{{\bf P}} 
\def\bS{{\bf S}} 
\def\bq{{\bf q}}

\newcommand{\todo}[1]{}
\renewcommand{\todo}[1]{\textcolor{blue}{\small\texttt{[:~#1~:]}}}

\title{On the energy balance of Newtonian Gravitation}

\author{R. Trasarti-Battistoni$^1$, Máté Pszota$^{2,3, ^{*}}$  and P. V\'an$^{2,4}$ }
\address{$^1$ Independent researcher, Brescia, Italy;\\
$^2$Department of Theoretical Physics, HUN-REN Wigner Research Centre for Physics, H-1525 Budapest, Konkoly Thege Miklós u. 29-33., Hungary; \\
$^3$Institute of Physics and Astronomy, Faculty of Science, Eötvös Loránd University, H-1053 Budapest, Pázmány Péter stny. 1/A, Hungary; \\
and  $^4$Department of Energy Engineering, Faculty of Mechanical Engineering,  Budapest University of Technology and Economics, H-1111 Budapest, Műegyetem rkp. 3., Hungary}
\date{\today}

\begin{document}
\begin{abstract}
It is shown that the energy density formulas of Newtonian gravity by Maxwell, Bondi and Ohanian cannot be distinguished by boundary conditions, and also the corresponding energy balances are identical. However, they are not equivalent. From a thermodynamic point of view, the Ohanian energy density is distinguished. 
\end{abstract}

\maketitle

\begingroup
\renewcommand{\thefootnote}{\fnsymbol{footnote}}
\footnotetext[1]{Corresponding author, \texttt{pszota.mate@wigner.hun-ren.hu}}
\endgroup

\section{Introduction}

The energy of the gravitational field is not uniquely defined. Several expressions for energy density and energy flux have been proposed, and this ambiguity is usually attributed to boundary conditions at spatial infinity \cite{Pet81a}. 

From a physical point of view, however, assumptions about spatial infinity are unsatisfactory. In this paper, we argue that the problem should instead be analyzed locally, using boundary conditions on energy fluxes rather than on energy densities. In this way, the issue can be reformulated entirely in terms of differential energy balances. Because reference frames play a crucial role in such a local treatment, we draw on concepts and methods from fluid mechanics. 

The paper is organised as follows. First, we review the basic concepts of energy and energy flux in Newtonian gravity (NG) and show that mass balance plays an essential role. We then introduce the necessary continuum-mechanical framework, in particular the distinction between local and substantial balances and between laboratory and comoving frames. Next, we discuss the energy balances associated with the Ohanian, Maxwell, and Bondi energy densities, and we demonstrate their equivalence starting from the Ohanian form. Here, the concept of {classical holography} is illuminating. The concept of classical holography is central to this analysis: it allows the different balances to be transformed into one another and their source terms to be expressed either in bulk-force or in surface-force form. In Section 5, we show that the differences among these balances are governed by an observer-dependent, sourceless balance, the \textit{Green balance}. Finally, we discuss the implications of these results.

The paper also includes several appendices. Appendix \re{AppBalEnergy} discusses the origin of the Ohanian energy balance from a general perspective and emphasizes an energy-based argument in comparison to the original thermodynamic one.
Appendix \re{AppBalq} presents further implications and properties of the Green balance. Appendix \re{AppLagrangian} analyses gravitational energy density from the point of view of the Lagrangian theory of classical fields.

\section{Newtonian Gravity and Continuum Description}

For notational convenience, we adopt units such that $4\pi G=1$, where $G$ is Newton's universal gravitational constant. We use the standard shorthand notation 
$\pt{}\equiv\frac{\partial}{\partial t}$ for the partial (local, Eulerian) time derivative and $\nabla\equiv\frac{\partial}{\partial \bx}$ for the usual spatial gradient in Cartesian coordinates. The total (substantial, convective or Lagrangian) time derivative is denoted by $^\cdot\equiv\frac{d}{dt} \equiv \pt{} + \bv\cdot\nabla$. 
In particular, for an arbitrary scalar field $\phi(\bx,t)$, its spatial gradient is the vector $\bV:=\nabla\phi$,
and $(\nabla\phi)^\cdot\equiv\frac{d\bV}{dt}$ denotes the substantial time derivative applied to the entire vector field.

Consistently with Newtonian gravity, the background geometry is assumed to be Euclidean.

\subsection{Fields and fluids} 
We make the following assumptions:

\begin{enumerate}

\item We assume the fields are regular at each position $\bx$ and time $t$, so that they are sufficiently differentiable. Some fields are also assumed to vanish sufficiently rapidly at spatial infinity. The material system is modeled as a continuous (one-component) fluid characterized by a mass density distribution field $\rho(\bx,t)$, a velocity field $\bv(\bx,t)$, and an associated convective mass current density (or momentum density) $\bJ_v(\bx,t):=\rho\bv$. 
All gravitational effects are fully described by the gravitational field strength $\bg(\bx,t)$, and the associated gravitational scalar potential field $\varphi(\bx,t)$.

\item The essential requirement is the existence and near-uniqueness of the  gravitational scalar potential $\varphi(\bx,t)$, defined such that it is related to the gravitational vector strength field $\bg(\bx,t)$ by:
\eqn{g=-gradphi}{
\bg(\bx,t) =:-\nabla \varphi(\bx,t)
}
Here, "near-uniqueness" means uniqueness up to a largely but not completely irrelevant spatially uniform offset $\varphi_*(t)$, which may vary in time. This represents a simple gauge freedom, or a "gauge ambiguity".

\item The time evolution of the fluid is governed by two equations of motion: 

(i) The evolution of $\rho(\bx,t)$ is governed by the mass  conservation equation:
\eqn{mass-flow-cont}{
\pt{\rho}+ \nabla\cdot(\rho\bv) =0
}

(ii) The evolution of $\bv(\bx,t)$ is governed by the momentum balance (Cauchy equation):
\eqn{momentum-force}{
\rho\dot\bv + \nabla\cdot\bP
= \rho\left( \pt{\bv} + \bv\cdot\nabla\bv \right)+ \nabla\cdot\bP
= \mathbf{f} 
:= \left( \mathbf{f}_g+\mathbf{f}_{other} \right)
}
where the pressure force density is $\mathbf{f}_p=-\nabla\cdot\bP$, and the total force density is $\mathbf{f} = \mathbf{f}_g + \mathbf{f}_{other}$, with gravitational contribution $\mathbf{f}_g=\rho\bg$. In what follows, we assume $\mathbf{f}_{other}\equiv0$. 

\item The spatial distribution of the gravitational fields is determined by two equations:

(i) The intuitive picture that field lines of gravity $\bg=-\nabla\varphi$ 
are "sourced" by (more precisely: "sink" into) the mass density distribution $\rho$,
and the way $\bg$ "spreads" all around $\rho$ according to Newton's inverse square law
is formally described by 
the NG Gauss theorem (in terms of $\bg$) or the Newton-Gauss-Poisson equation (in terms of $\varphi)$,
defining the Laplacian operator $\Delta:=\nabla\cdot\nabla$:
\eqn{Gauss-Poisson}{
\Delta\varphi=-\nabla\cdot\bg=\rho}
(ii) The intuitive picture that field lines of $\bg$ neither "turn around" nor "close upon themselves"
is formally described by the NG "irrotationality/conservativity condition":
\eqn{irrotcons}{
\nabla\times\nabla\varphi=-\nabla\times\bg \equiv0 
}
\end{enumerate}

Remarkably, within NG (\textit{i.e.} assuming equation \re{Gauss-Poisson}), the gravitational force density $\mathbf{f}_g$ can be written in divergence form, analogously to the pressure force density $\mathbf{f}_p=-\nabla\cdot\bP$: 
\eqn{f_g=rhog}{
\mathbf{f}_g
=\rho\bg
= - \nabla\cdot\bP_g 
}
Here $\bP_g$ is a gravitational pressure tensor, analogous to the Maxwell stress tensor in electromagnetism. The property described by the relation in eq. \re{f_g=rhog} is referred to as the \emph{classical holographic property}, and its importance is discussed in detail in Section \ref{classical_holography}.

Both \re{Gauss-Poisson} and \re{irrotcons} \emph{for the fields $\bg$ or $\varphi$ themselves}  are well-known equations of NG.  Considerably less well-known are the following relations, that involve  their \emph{partial time derivatives $\pt{\bg}$ and $\pt{\varphi}$}.

First of all, the \emph{gravitational "displacement current"} field is defined as:
\eqn{gravdisplcurr}{
\bJ_g(\bx,t):=-\pt{\bg}(\bx,t)
}
It is (with the opposite sign) the NG analogue of Maxwell's electric "displacement current" $\bJ_E:=\epsilon_0\pt{\bE}$ of EM, proposed by Maxwell himself 
\cite{Max1865viii,Max1873-79,Roche98, Kitano24}.

Although $\bJ_g=-\pt{\bg}$, is not a mass current, it has the same physical dimensions as the convective mass current $\bJ_v=\rho\bv$,  which motivates the term "current" and the notation $\bJ$. 
The word "displacement" goes back to an obsolete theoretical model originally used by Maxwell. 

Considering $-\pt{\bg}$, taking its divergence, exchanging derivatives $\nabla\cdot$ and $\pt{}$ yields:
\eqn{-divptg=mass-flow-cont}{
\nabla \cdot \left(-\pt{\bg}\right) = \pt{\rho} = -\nabla\cdot(\rho\bv)
}
where the l.h.s. equality comes from the NG Poisson equation \re{Gauss-Poisson}, while the r.h.s. equality comes from mass-flow continuity \re{mass-flow-cont}. Relation \re{-divptg=mass-flow-cont} plays a key role in the rest of the paper, 
and it could be referred to as the "displacement-continuity" equation.

\section{NG energy densities, energy current densities, and balances}

This Section summarizes, in a systematic way, results previously obtained or used in Synge \cite{Syn72a}, Peters \cite{Pet81a}, Sebens \cite{Sebens2017-lu} and \cite{Sebens2022-bh}, Dewar \& Weatherall \cite{DewWea18a}, Duerr \& Read \cite{DueRea19a}, Vasyliunas \cite{Vas22a}, Eklund \cite{Ekl22a}, Bengtsson \& Eklund \cite{BenEkl23a}. 
Each of these works considered some of the cases dealt with here, 
but only \cite{Sebens2022-bh}, \cite{Ekl22a}, and \cite{BenEkl23a} treated all the cases together.
Altogether, the pairs of energy density -- energy current density referred here below as 
Maxwell-Livens and Bondi-Bondi  have received, so far, comparably more attention than the case Ohanian-No-Name, which is instead somewhat favoured by the present paper. 

Energy balance, in general:
\eqn{eps_a_balance}{
\pt{\eps_a} + \nabla\cdot\bS_a = w}
The first term on the l.h.s. is the time variation of the local energy density $\eps_a$, and the second term on the l.h.s. is the spatial divergence of the corresponding local energy current density $\bS_a$.
Here the index $a$ conveniently parametrizes various energy density / energy current density pairs \cite{BenEkl23a}. 
Other Authors used similar, mathematically equivalent parametrizations earlier on,
but the parametrization of \cite{BenEkl23a} is the most convenient here.

The precise mathematical role of the parameter $a$ will become clear in Sect.\re{lincomb}. For the moment, one can regard the value of $a$ just as a useful mnemonic device: $a=1$ for Maxwell; $a=0$ for Ohanian; $a=1/2$ for Bondi; and just $a$ for "arbitrary".


The source term on the r.h.s. is the power density $w= -\mathbf{f}_f\cdot\bv$ delivered to the matter by the field force density $\mathbf{f}_f$ acting upon it. 
It is the NG analogue of Joule's power, delivered from the electric field to the current, in EM.
This is the usual point of view. 

Alternatively \cite{Sebens2017-lu}, one might define the opposite $+\mathbf{f}_m=-\mathbf{f}_f$ 
as the force exerted by matter upon the field, 
and correspondingly the power $+\mathbf{f}_m\cdot\bv = -\mathbf{f}_f\cdot\bv$
delivered to the field by the matter force density $\mathbf{f}_m$ acting upon it.
Still alternatively, if the force can be re-expressed as $\mathbf{f}_f=-\nabla\cdot\bP_f$, then the power can be re-expressed in terms of tensors $[\bP]:[\nabla\bv]$ instead of in terms of vectors as $(\nabla\cdot\bP_f)\cdot\bv$. However, fixing the power density on the r.h.s. of \re{eps_a_balance} does not uniquely fix either of the two terms on the l.h.s. separately, but only their sum.

\subsection{NG energy density $\varepsilon(\bx,t)$}

In fact, there is an infinite number of energy densities $\eps_a(\bx,t)$
(a continuous 1-parameter family described by $a$), which are locally different, but globally equivalent: 
their distributions are different point-by-point in space,
but when integrated over all space \emph{under suitable boundary conditions at spatial infinity}
(vanishing surface integrals), 
and \emph{assuming the Poisson equation} ("to turn $\varphi$ into $\rho$", and \textit{vice versa}, so to say),
they return the same total gravitational energy $E_g$, even for ${a'}\not={a"}$:
\eqn{totE_g}{
\int d^3\bx \eps_{a'}
= E_g =
\int d^3\bx \eps_{a"}
}

Remarkably interesting, for various reasons, are three energy densities $\eps_1$, $\eps_0$, $\eps_{\frac{1}{2}}$.
For clarity, they are all listed and defined, with their characteristic properties, here:
\\

\begin{enumerate}
 
\item ($a=1$) Maxwell \cite{Max1865viii}, pure \emph{field} energy density (negative):
\eqn{eps_1}{
\eps_{Maxwell} \equiv 
\eps_1:= -\frac{1}{2}\bg^2 
= -\frac{1}{2}(\nabla\varphi)^2
\leq 0
}
Expression \re{eps_1} was originally suggested by Maxwell himself,
in the \emph{Note on the attraction of gravity} 
at the end of his paper \emph{A dynamical theory of the electromagnetic field} \cite{Max1865viii}.
Maxwell clearly declared himself puzzled because $\eps_1$ is negative, 
and he was not able to find a physical interpretation that was convincing to him. 
\\

\item ($a=0$) Ohanian \cite{Ohanian2013-ez}. It is a matter-energy \emph{interaction}
+ pure \emph{field} energy density:
\eqn{eps_0}{
\eps_{Ohanian} \equiv
\eps_0:= +\rho\varphi +\frac{1}{2}\bg^2 
= \rho\varphi + \frac{1}{2}(\nabla\varphi)^2
}
Both terms are interpreted as associated with gravitation, either through $\varphi$ or $\bg$.
The first term $+\rho\varphi\leq0$ can be interpreted as matter-field inter-action 
(\emph{not potential} energy $\frac{1}{2}\rho\varphi$ \textit{i.e.}
\emph{not} mutual matter-matter interaction energy,
no prefactor $1/2$ to avoid double counting of particle pairs),
the second term $+\frac{1}{2}(\nabla\varphi)^2>0$ as (positive) energy density 
purely due to the gravitational field $\bg=-\nabla\varphi$.

Remarkably, Ohanian's energy density \re{eps_0} equals (the opposite of)
$\mathcal{L}_g$, the well-known "gravitational" Lagrangian density
commonly used to derive the field equations \re{Gauss-Poisson}, \re{irrotcons} of Newtonian Gravitation
from the point of view of the Lagrangian theory of classical fields, e.g.\cite{Fra15a,DewWea18a}.
It was then further modified by Eklund \cite{Ekl22a} and Bengtsson \& Eklund \cite{BenEkl23a},
to derive also the equations of motions \re{mass-flow-cont}, \re{momentum-force} of the matter fields.
For details see Appendix \re{AppLagrangian}.

Finally, \re{eps_0} is the same starting expression assumed by Ván and Abe \cite{VanAbe22a}
in their thermodynamic framework for gravity.
For details see Appendix \re{AppBalEnergy}.
\\

\item ($a=1/2$) Bondi \cite{bondi1965}  \emph{potential} energy density 
(=$1/2$ matter-field interaction, negative for $\varphi\to0$ at spatial infinity). It is the arithmetical mean of the energy densities \re{eps_1} and \re{eps_0}:
\eqn{eps_1/2}{
\eps_{Bondi} \equiv 
\eps_{\frac{1}{2}}:= +\frac{1}{2}\rho\varphi
= \frac{1}{2}\varphi\Delta\varphi
\leq 0
}
It is the gravitational interaction energy of particles. The prefactor $1/2$ avoids double-counting the interactions.
\end{enumerate}

\subsection{NG energy current densities $\bS(\bx,t)$}

Now, fixing both the power on the r.h.s. of \re{eps_a_balance} 
and one at will of the two terms on its r.h.s.
does uniquely fix the remaining third term on the l.h.s.
However, fixing $\eps_a$ as above, and hence $\pt{\eps_a}$ as a consequence,
only fixes $\nabla\cdot\bS_a$ but not $\bS_a$ itself.
As for energy densities $\eps_a$, there is an infinite number of energy currents $\bS_a(\bx,t)$
locally different, but globally equivalent. 
But the ambiguity for energy currents is even wider than for energy densities themselves:
for a given $a$ hence given $\eps_a$ and $\pt{\eps_a}$, there can be two (more precisely, infinite) 
locally different energy currents $\bS'_a\not=\bS^{"}_a=\bS'_a+\nabla\times\bW$, 
with an arbitrary vector (potential) field $\bW(\bx,t)$,
which nevertheless share the same divergence $\nabla\cdot\bS_a'=\nabla\cdot\bS_a^{"}$, 
since $\nabla\cdot(\nabla\times)\equiv0$.
A similar problem is encountered in the context of classical Electro-Magnetism, 
where it is widely discussed, see e.g. \cite{Sle42}, \cite{KTMcD20alt}, \cite{MorLej22}, and references therein.

Remarkably interesting, for various reasons, 
are the three energy current densities $\bS_1$, $\bS_0$, $\bS_{\frac{1}{2}}$,
which in some sense most directly correspond to 
the three energy densities $\eps_1$, $\eps_0$, $\eps_{\frac{1}{2}}$ defined above.
Each of them contains two parts: 
one related to mass motion (advection), $\bS_v:=(\rho\bv)\varphi$,
like in electric engineering \cite{MorLej22},
and another related to the time variation of gravitation.
In turn, there are two alternative gravity-related energy currents:
$\bS_g := (-\pt{\bg})\varphi$ (Synge \cite{Syn72a} eq.(5.11)), 
proportional to the scalar potential
and parallel (assuming $\varphi<0$) to the vector field time-variation, 
or instead 
$\bS_\varphi := \bg(\pt{\varphi})$ (Bengtsson \& Eklund \cite{BenEkl23a} eq.(51)),
proportional to the scalar potential time-variation, 
and parallel to the vector field.

The three energy current densities are:
\\

\begin{enumerate}

\item $(a=1)$, Livens-like \cite{Livens1917-us},
Vasyliunas \cite{Vas22a} eq.(20), 
scalar potential advection by the total (displacement + mass) current:
\eqn{S_1}{
\bS_1
= (\bS_g+\bS_v)
= (-\pt{\bg})\varphi  +\rho\bv\varphi
= (+\pt{\nabla\varphi})\varphi + \rho\bv\varphi
}

\item $(a=0)$ "No-Name" of its own,
potential variation $\times$ gravitational field
+ scalar potential advection by mass current:
\eqn{S_0}{
\bS_0 
= (\bS_\varphi + \bS_v)
= +\bg(\pt{\varphi})             + \rho\bv\varphi
= (-\nabla\varphi)(\pt{\varphi}) + \rho\bv\varphi
} 

\item $(a=1/2)$ Vasyliunas \cite{Vas22a} eq.(34),
arithmetical mean of the energy current densities \re{S_1} and \re{S_0},
including scalar potential advection by mass current:
\eqn{S_1/2}{
\bS_{\frac{1}{2}} = \frac{(\bS_1+\bS_0)}{2}
= \frac{1}{2}\left[ -(\pt{\bg})\varphi + \bg(\pt{\varphi}) \right] 
+ \rho\bv\varphi
= \frac{1}{2}\left[ (\pt{\nabla\varphi})\varphi -(\nabla\varphi)(\pt{\varphi}) \right]
+ \rho\bv\varphi
}
The expression in square brackets is the arithmetic mean of $\bS_g$ and $\bS_\varphi$, 
which Synge \cite{Syn72a} Sect.8 interprets as
"equal sharing of interaction energy between two interacting subsystems".
Without mass advection term $\rho\bv\varphi$
it reduces to Synge's eq.(8.1), who considered $\rho=0$.

\end{enumerate}

Note that the purely advective energy current $\rho\bv\varphi$ appears intact in all three energy currents $\bS_1$, $\bS_0$, and $\bS_{\frac{1}{2}}$. Note also that the two gravity-related terms in (22) handle the gauge ambiguity of the spatial offset $\varphi_{*}(t)$ differently. It might initially appear that in $\bS_\varphi=(-\partial_{t}\varphi)(\nabla\varphi)$, the ambiguity is removed because the derivatives act upon distinct factors. However, if the arbitrary offset varies in time ($\dot{\varphi}_{*}(t) \neq 0$), then the partial time derivative shifts as $\partial_{t}\tilde{\varphi} = \partial_{t}\varphi + \dot{\varphi}_{*}$. Therefore, $\bS_\varphi$ also perpetuates the offset ambiguity, just as $\bS_g=+\varphi(\partial_{t}\nabla\varphi)$ does. Remarkably, despite the gauge-dependence of these individual currents, this ambiguity perfectly and rigorously cancels out in every local energy balance \re{enbal_1}-\re{enbal_1/2}, guaranteeing the physical consistency of the overall framework.

\subsection{Energy balances}
\label{sect:enbal}

Substituting each pair $\eps_a,\bS_a$ in \re{eps_a_balance} gives a corresponding energy balance,
as in \cite{Sebens2022-bh}, \cite{Ekl22a}, and \cite{BenEkl23a}.

All the energy balances in the form above for the cases $a=1,0,1/2$ are explicitly listed here, with their characteristic properties, pro's and con's:

\begin{enumerate}
\item $(a=1)$ Maxwell-Livens;
Maxwell's gravitational field energy density $\eps_1$ has a very simple mathematical expression, Maxwell's displacement current in $\bS_1$, not quite such a simple expression;
rather well-established physical interpretations for both, yet somewhat obscure;
favored by analogy of NG with EM:
\eqn{enbal_1}{
\pt{} \left( -\frac{1}{2}\bg^2 \right)
+\nabla\cdot \left[ -(\pt{\bg})\varphi  +(\rho\bv)\varphi \right]
= -\rho\bv\cdot\bg
}
\\

\item $(a=0)$ Ohanian-No-Name;
not too simple mathematical expressions of $\eps_0$ and $\bS_0$; somewhat unclear physical picture of both;
but favored by different arguments from  Classical Lagrangian Field Theory App.\re{AppLagrangian}, 
and Thermo-Dynamics of continuous media App.\re{AppBalEnergy}:
\eqn{enbal_0}{
\pt{}\left( +\rho\varphi+\frac{1}{2}\bg^2 \right)
+\nabla\cdot \left[ +\bg(\pt{\varphi})  +(\rho\bv)\varphi \right] 
= -\rho\bv\cdot\bg
}
\\

\item $(a=1/2)$ Bondi-Bondi;
Bondi's gravitational potential energy of particles $\eps_{\frac{1}{2}}$, 
with a simple expression and well-established physical interpretation in NG; 
all the contrary for Bondi's current $\bS_{\frac{1}{2}}$,
awkward mathematical expression, and unclear physical interpretation;
arithmetical mean of the energy balances \re{enbal_1} and \re{enbal_0}:
\eqn{enbal_1/2}{
\pt{}\left( +\frac{1}{2}\rho \varphi \right)
+\nabla\cdot \left[ +\frac{1}{2}\bg(\pt{\varphi}) - \frac{1}{2}(\pt{\bg})\varphi 
+(\rho\bv)\varphi \right] 
= -\rho\bv\cdot\bg
}
\\
This expression can be extracted from the energy balances in the standard textbook
Galactic Dynamics by Binney \& Tremaine \cite{BinTreBook}, appendix F.1.3.
\end{enumerate}

Note that, while the power source on the r.h.s. is the same for all three cases (and in general, arbitrary $a$),
the energy densities and energy current densities on the l.h.s.,
while having clearly different mathematical expressions, have also markedly different physical distributions in space.
For example, this is vivedly illustrated by 
Fig.1 in \cite{Sebens2022} for $\eps_\frac{1}{2}$ vs $\eps_1$, in the case of a spherical ball of uniform mass density,
and Fig.2.6 in \cite{Ekl22a} for $\bS_\frac{1}{2}$ vs $\bS_1$ vs $\bS_0$,
in the case of two equal masses orbiting around each other.

\subsection{The	gravitational Poynting vector}
\label{sect:enbal_x}

{\color{black}
Remarkably, the case $a=1$ can be rewritten in a very suggestive, alternative way, closely analogous to its EM counterpart.

First of all, as originally noted by Heaviside \cite{Hea1893a,Hea1893b},
there are many and strong analogies between NG and classical (Maxwell+Lorentz) EM, and/or 
with Gravito-Electro-Magnetism (GEM), see e.g. Sebens \cite{Sebens2022}, Hermann \& Polig \cite{HermannPohlig22}.
However, 
NG is a fully non-relativistic theory (instantaneous speed of propagation of physical effects, action-at-distance),
while GEM like EM is a fully relativistic theory (finite speed of propagation of physical effects, retardation.)
So it is more appropriate to compare NG with \emph{quasi-static} sub-models of full dynamic EM, 
in particular the Electro-Quasi-Static (EQS) model, discussed e.g. by Larsson \cite{Larsson07}. 

As a consequence of such analogies, 
one can define the \emph{gravito-magnetic} field $\bb$
(sometimes called "gravinetic" field \cite{KruBed87a}), 
the NG analog of the magnetic field of EM.
This has been considered by many authors, from several different theoretical points of view,
leading to slightly different versions of the NG vs EM analogy, e.g.: 
Heaviside \cite{Hea1893a}, \cite{Hea1893b} from NG in analogy with EM,
Bedford \& Krumm \cite{BedKru85a} from NG plus SR,
Kolbenstvedt \cite{Kol88a} from SR plus a little bit of GR (gravitational time dilation),
Harris \cite{Har91a} or Ohanian \& Ruffini \cite{Ohanian2013-ez},
from GR plus linear approximations (weak field, slow motions), 
Buchert \cite{Buc06a} or Vasyliunas \cite{Vas22a} purely from NG, and so on.
Note that units, signs, and notation can vary widely amongst different works,
even worse than in usual EM. 
In particular, there are different factors of $2$ or $4$ 
in the various definitions of $\bb$ and/or of the gravitational Lorentz force,
that in some cases give different physical predictions, not just different mathematical definitions.
(And in some versions, e.g. \cite{Har91a} there are additional terms, 
with no strict EM analogues, due to residual non-linearities inherited from GR.)

In all versions of the NG-EM analogy, the gravito-magnetic field $\bb$ satisfies two field equations, 
the NG analogue of the Ampére-Maxwell equation:
\eqn{Ampere-Heaviside}{
\nabla\times\bb = -\rho\bv + \pt{\bg}
}
and the NG analogue of the magnetic Gauss theorem ("no magnetic poles"):
\eqn{Gauss-Thomson}{
\nabla\cdot\bb = 0
}
Note that the "obvious" NG analogue of the Faraday's induction equation would be, but is not:
\eqn{Faraday-Heaviside}{
\nabla\times\bg = -\pt{\bb}.
}
This equation holds in GEM and EM (with $\pt$ replaced by $\frac{1}{c}\pt{}$),
but it \emph{does not} hold in strict NG and EQS (and furthermore, there is no $c$ in NG):
it might well happen that $\pt{\bb}\not=0$ at some instant and position,
but for sure it will be $\nabla\times\bg\equiv0$ always and everywhere, because of the field equation \re{irrotcons}.
In contrast with GEM or EM, in NG and EQS the (gravito)magnetic field is just an "auxiliary" field, 
"accompanying" the (gravito)electric field in its time-evolution, 
but it pays no dynamic role on its own: 
because of \re{irrotcons} there is no gravitational "Faraday" induced electric field and force, 
because of \re{f_g=rhog} there is no gravitational "magnetic Lorentz-Laplace" velocity dependent force, 
and because of both (non-existing magnetic-related forces make no "magnetic" work)
there is no gravitational "magnetic" energy density in $\eps_a$,
only physical quantities related to the gravito-"electric" field $\bg$:
itself, its source $\rho$, and its potential $\varphi$.
%
Furthermore, one can define the \emph{gravitational Poynting vector} $\bS_\times:=-\bg\times\bb$,
the NG analog of the well-known Poynting's vector of EM \cite{Poy1884}.
This has been discussed by several Authors, 
e.g. \cite{KruBed87a}, \cite{Hni97a}, \cite{Vas22a}.
%
Finally, see in particular \cite{Vas22a}, it is possible to prove the NG analogue of the Poynting's Theorem
\cite{Poy1884}, \textit{i.e.}the following energy balance:

$(a=1)$bis: Maxwell-Poynting; same simple mathematical expression for $\eps_1$; 
even simpler mathematical expression for $\bS_\times$;
rather well-established physical interpretations for both, yet somewhat obscure; 
favored by analogy of NG with EM:
\eqn{enbal_1x}{
\pt{}\left( -\frac{1}{2}\bg^2 \right)
+\nabla\cdot \left[ -\bg\times\bb \right]
= -\rho\bv\cdot\bg
}
} 
Again, note that, while the power source on the r.h.s. and even the energy density on the r.h.s.
are the same in \re{enbal_1} and \re{enbal_1x}, yet the spatial distributions of $\bS_1$ vs $\bS_\times$
are very different. In the analogous EM case this is well-known and studied.
For visual illustrations of $\bS_\times$ in the case of simple electric circuits, 
and how $\bS_\times \not = \rho\bv\varphi$ 
(i.e. electric energy in a circuit does not flow along the wires, 
but through space around it, via the electromagnetic field)
see e.g. \cite{MorSty12}, and references therein.

\subsection{Linear combinations of the energy balances}
\label{lincomb}


{\color{black}
Note that each of the three energy densities \re{eps_1}, \re{eps_0}, \re{eps_1/2},
can be written as a simple linear combination of the other two: 
Maxwell's $\eps_1=2\eps_{\frac{1}{2}}-\eps_0$ while
Ohanian's $\eps_0=2\eps_{\frac{1}{2}}-\eps_1$, 
\textit{i.e.}each is twice Bondi's energy density minus the other one,
and in particular Bondi's $\eps_{\frac{1}{2}}=\frac{1}{2}\eps_0+\frac{1}{2}\eps_1$
is just halfway between Maxwell's and Ohanian's energy densities.
%
{\color{black}
As originally pointed out by Peters \cite{Pet81a},
in principle, one could choose any two equivalent energy densities $\eps_a'$ and $\eps_a''$
(\textit{i.e.}, yielding the same integrated total gravitational energy $E_g$, as in \re{totE_g}), 
and linearly combine them to get any other arbitrary (but still equivalent) energy density $\eps_a$.
In practice, it turns out particularly convenient to parametrize all energy densities as follows:
\eqn{eps_a_lincomb}{
\eps_a = a\eps_1 + (1-a)\eps_0
}
choosing Maxwell's $\eps_1$ and Ohanian's $\eps_0$ energy densities as the two building blocks.
However, these linear combinations are not the only possibilities, 
there are other eligible building blocks, see the Green balance \re{ptPoisson_etc} later on.
}

Note that the parametrization \re{eps_a_lincomb} is an \emph{on-shell} relation: 
it assumes the Poisson equation \re{Gauss-Poisson}, 
which allows  to replace $\rho\varphi$ by $\varphi\Delta\varphi$ (and vice versa). 
Off-shell, \textit{i.e.}\ without assuming the field equation, 
the energy densities $\eps_0$ and $\eps_1$ are structurally different: 
$\eps_0:=\frac{1}{2}\rho\varphi$ contains the coupling of matter $\rho$ with (potential) field $\varphi$, 
whereas $\eps_1:=-\frac{1}{2}\bg^2$ is a pure (strength) field expression.
} 


As previously done for energy densities in \re{eps_a_lincomb},
it turns out particularly convenient to parametrize their corresponding energy current densities as follows:
\eqn{S_a_lincomb}{
\bS_a = a\bS_1 + (1-a)\bS_0
}
choosing Livens' $\bS_1$ and "No-Name"'s $\bS_0$ energy current densities as the two building blocks.
Note that \re{eps_a_lincomb} and \re{S_a_lincomb} use the same linear combination for
energy densities and energy current densities, \textit{i.e.}the same weights $a,(1-a)$ for corresponding $\eps,\bS$ pairs.
The distinction about off-shell and on-shell situations, already seen for the $\eps$'s, apply to the $\bS$'s as well.

As previously done for the energy densities in \re{eps_a_lincomb}, 
and for the energy current densities in \re{S_a_lincomb},
it turns out to view the energy balances themselves, 
for each arbitrary value of the parameter $a$, as the corresponding linear combinations of 
the Maxwell-Livens' balance \re{enbal_1} and the Ohanian-No-Name's balance \re{enbal_0},
with corresponding weights $a,(1-a)$, as before.
In particular, the Bondi-Bondi balance 
eq.\re{enbal_1/2} = $\frac{1}{2}$ eq.\re{enbal_1} + $(1-\frac{1}{2})$ eq.\re{enbal_0}.
In general, for arbitrary $a$ one gets \re{eps_a_balance}.

Summarizing the discussion so far:
\begin{enumerate}

\item There are several proposed, competing pair candidates 
of energy density $\eps_a \leftrightarrow \bS_a$ energy current density
(indeed there are even more, due to the inherent ambiguity of the current inside the divergence); 
they can be conveniently parametrized with a single real number $a$.

\item Each pair satisfies an energy balance (NG analogue of Poynting's theorem of EM), with the same source (the NG analogue of the Joule power of EM). Is there a "best" pair $\eps_a \leftrightarrow \bS_a$?
Particularly interesting candidates are 
the Maxwell-Livens's ($a=1$, also Maxwell-Poynting), 
the Ohanian-No-Name's ($a=0$), and the Bondi-Bondi ($a=1/2$).

\item In any case, the volume-integrated gravitational energy is not enough to discriminate among different energy densities. A straightforward idea proposed by several authors is to investigate the role of boundary conditions, that constrain the surface integral, as seen above.

\item The definitions of the various energy current densities 
tend to mix up the role played by Newtonian Gravitation (and its time variation) with those of fluid mechanics (mass motion). 

\end{enumerate}

Therefore, in the next sections, the required concepts are briefly introduced while focusing on two particular aspects of non-relativistic continua. 
The first one is the transformation between customary reference frames: 
the "fixed" laboratory frame and the frame comoving with the fluid,
called Eulerian (partial time derivatives) and Lagrangian (substantial time derivative) in fluid mechanics.
The other one is a recently observed general property of ideal fluids and scalar fields:
the classical holography  \cite{Van23a,SzucsVan25m}. 

\section{Continuum foundations}

\subsection{Balances} The balances of continuum mechanics and nonequilibrium thermodynamics express the spacetime properties of extensive physical quantities. A local balance of a scalar $A$ can be written as
\eqn{a_locbal}{
\pt{(\rho \alpha)} + \nabla\cdot \bJ_\alpha = \sigma_\alpha. 
}
Here $\rho \alpha$ is the density of $A$ and $\alpha$ is the (mass)specific $A$. $\bJ_\alpha$ is the total current density of $A$ and $\sigma_\alpha$ is the production density rate of $A$. $\pt{ }$ is the partial time derivative and $\nabla\cdot$ is the divergence. One can obtain the substantial form of the balance by introducing the substantial time derivative, which for $\alpha$ reads $\dot \alpha= \partial_t \alpha + \bv\cdot\nabla \alpha$. Then  
\eqn{a_presubbal}{
(\rho \alpha\dot) + \nabla\cdot \left( \bJ_\alpha - \rho \alpha \bv  \right)
+ (\rho \alpha) \nabla\cdot \bv = \sigma_\alpha,
}
and $\rho \alpha\bv$ is the advective  current density, 
while $\bj_\alpha := (\bJ_\alpha - \rho \alpha\bv)$ is the conductive current density of $A$. 

If $\alpha=1$, one gets the mass balance. Mass is conserved, therefore $\sigma_m = 0$;
and the mass current density has only an advective part $\bJ_v=\rho\bv$, 
so the conductive part of the current density of mass is zero, $\bj_m:=(\bJ_m - \rho\bv_m)=0$.
Therefore, the local and substantial forms of the mass balance are:
\eqn{massbalance}{
\pt{\rho} + \nabla\cdot(\rho\bv) = 0, \qquad 
\dot\rho + \rho\nabla\cdot{\bv} =0.
}

Then the general form of substantial balances, that is  \re{a_presubbal} simplifies to
\eqn{a_subbal}{
\rho \dot \alpha + \nabla\cdot \bj_\alpha  = \sigma_\alpha.
}
The difference of local and substantial balances is related to changes of reference frames. The local balance is the form from the point of view of an external laboratory frame. The substantial balance is the form obtained when the reference frame is fixed to the continuum itself. Naturally, it requires that the defining fields of the continuum must have some regularity requirements. Moreover, in a substantial balance the comoving form is expressed with the physical quantities relative to the laboratory frame (e.g. the relative velocity field  $\bv$ is the same). The most important background knowledge is that the balance itself is observer independent: it is a spacetime divergence of a four-quantity. This is an evident fact in special relativity but it is the same in Galilean relativistic (nonrelativistic) spacetime as well (see e.g. \cite{Mat20b}). 
The local balance is a laboratory frame version of the covariant law, the substantial balance is the same in a comoving frame, but expressed with the relative quantities of the labframe. The substantial time derivative is comoving with the fluid.

It is remarkable, that the substantial time derivative and the spatial derivative $\nabla$ are not commutative operations:
\eqn{stder}{
(\nabla\varphi\dot) = \nabla(\dot\varphi) -\nabla\varphi\cdot\nabla\bv &,
\qquad 
(\partial_i\varphi\dot) = \partial_i(\dot\varphi) -\partial_k\varphi\partial_iv^k.
}

\subsection{Classical holography}
\label{classical_holography} It is also remarkable that the bulk force due to $\mathbf{f_g}=\rho\bg$ of the gravitational field $\bg=-\nabla\varphi$ in any spatial region can be expressed as a surface traction on the boundary, because it can be written as a divergence of a related gravitational pressure tensor. This is called the \emph{classical holographic property} of the gravitational field.
In terms of gravitational pressure tensor ${\bf P}_{grav}$ and force density ${\bf f}_g$:
\eqn{chol_f}{
\nabla\cdot {\bf P}_{grav} = {\bf f}_g.
}
Expressing the gravitational Maxwell stress tensor in terms of the field strength $\bg$:
\eqn{chol_g}{
\nabla\cdot\left[\frac{1}{2} \bg^2  \bId -\bg\circ\bg \right] = +\rho\bg
}
or again, in terms of gravitational potential $\varphi$ and its gradient $\nabla\varphi$:
\eqn{chol_phi}{
\nabla\cdot\left[\frac{1}{2} (\nabla\varphi)^2  \bId -\nabla\varphi\circ\nabla\varphi \right] = -\Delta\varphi\nabla\varphi= -\rho\nabla\varphi.
}
The first equality is an identity, but the second one requires the validity of the Poisson equation.
While eq. \re{chol_f}  mathematically resembles the standard boundary-traction equivalence of a Cauchy stress tensor, it represents a deeper constraint, especially when viewed through the lens of weakly nonlocal thermodynamics as a constitutive property of the perfect fluid pressure tensor. In this framework, the energy balances discussed here are simplifications of a fundamental entropic formulation. For an ideal, dissipation-free continuous field, that is for perfect fluids, the vanishing bulk entropy production defines the pressure and the classical holographic property emerges as a consequence of zero dissipation  \cite{Van23a, SzucsVan25m}.

Consequently, the thermodynamic information required to specify the bulk equations of motion---specifically, the emergence of the Poisson equation---is entirely encoded by the boundary fluxes \cite{Van23a}. This reduction of bulk thermodynamic degrees of freedom to boundary information provides a non-relativistic, classical analogue to the holographic principle \cite{tHooft1993, Susskind1995}.

However, it is crucial to distinguish this classical thermodynamic holography from its quantum counterpart. While the thermodynamic state and macroscopic constraints are boundary-bound, the kinematic evolution of the fluid---the specific micro-arrangements of the density field $\rho(\bx,t)$ and momentum density field $\rho(\bx,t)\bv(\bx,t)$---remains strictly extensive. This macroscopic-microscopic duality mirrors Susskind's complexity argument \cite{Susskind2016}, which posits that while a system's entropy (information) is bounded by its boundary area, the computational complexity required to specify its exact internal dynamical state continues to scale linearly with the three-dimensional bulk volume. Thus, the classical weakly nonlocal fluid is thermodynamically holographic, yet kinematically volumetric.

\section{Gravitational energy balances: a continuum approach}


The evolution of any field quantity in continuum physics, therefore the corresponding  energy balance of any scalar field, is restricted by the Second Law of Thermodynamics. The obtained thermodynamic restrictions are particularly remarkable in the marginal case of ideal continua, with zero dissipation. If the entropy depends on the gradient of the field, then the second law restrictions lead to similar consequences as one could obtain from a Hamiltonian variational principle. In particular, for a gravitational field, if the local energy density of the continuum is the Ohanian one \re{eps_0}, then thermodynamic reasoning leads to the Poisson equation and also to the following substantial energy balance \cite{VanAbe22a,SzucsVan25m}:
\eqn{tgravbal_subst}{
\rho \left(\varphi+\frac{(\nabla\varphi)^2}{2\rho}\right)^{\bf \cdot}+
\nabla\cdot\left[-\dot\varphi \nabla\varphi \right] = 
 \left[\frac{1}{2}(\nabla\varphi)^2 \bId 
 -\nabla\varphi\circ\nabla\varphi \right]:\nabla\bv,
}
where the source term, the dissipation of gravitational energy, is a product of the gravitational pressure and the velocity gradient: typical for \emph{mechanical dissipation} in continuum mechanics. Thermodynamic derivations are based on material frames, therefore it is just natural to obtain a balance in a substantial form. It is also worth mentioning, that the conservation of mass \re{mass-flow-cont} is also a condition in the calculations. 

With the help of the {classical holographic property} \re{chol_phi} it can be transformed to the following form 
\eqn{tgravfbal_subst}{
\rho \left(\varphi+\frac{(\nabla\varphi)^2}{2\rho}\right)^\cdot +
\nabla\cdot\left[-\frac{(\nabla\varphi)^2}{2}\bv -\pt{\varphi}\nabla\varphi \right] =\rho\bv\cdot\nabla\varphi.
}
Here the identity $\nabla\cdot(\bv\cdot \bP_G) = \bv\cdot\nabla\cdot\bP_G + \bP_G:\nabla\bv$ is used.  remarkable that the source of the gravitational energy balance is the scalar product of the force density and the velocity, which is \emph{power density of the gravitational field}. 

There, one can see that the gravitational displacement vector, the negative partial time derivative of the gravitational potential appears in the energy current density. The transformation of the substantial derivatives to local ones leads to the local form of the energy balance 

\eqn{tgravfbal_loc}{
\partial_t\left(\rho\varphi
   + \frac{(\nabla\varphi)^2}{2}\right) + 
\nabla\cdot\left[\rho\bv\varphi - 
    \pt{\varphi }\nabla\varphi\right] = 
\rho\bv\cdot\nabla\varphi
}
Then, using the Green identity \re{ptLaplace_etc} and the balance of mass \re{massbalance}, one can get the balance of the gravitational energy with the Maxwell energy density alone:
\eqn{Mgravfbal_loc}{
\partial_t \left(-
    \frac{(\nabla\varphi)^2}{2}\right) + 
\nabla\cdot\left[\rho\bv\varphi+\varphi\pt{\nabla\varphi}\right] = 
\rho\bv\cdot\nabla\varphi
}
By transforming the partial time derivatives back to substantial ones one obtains:
\eqn{Mgravfbal_subst}{
\rho \left(-\frac{(\nabla\varphi)^2}{2\rho}\right)^\cdot +
\nabla\cdot\left[\varphi\pt{\nabla\varphi}+
\bv\frac{(\nabla\varphi)^2}{2} + \rho\bv\varphi \right] = \rho\bv\cdot\nabla\varphi
}
However, the complete substantial form requires eliminating the partial time derivatives of the laboratory frame. Therefore, using the {classical holographic property} of the gravitational pressure, the source term can be transformed back to the form of mechanical dissipation:
\eqn{MgravPbal_subst}{
\rho \left(-\frac{(\nabla\varphi)^2}{2\rho}\right)^\cdot +
\nabla\cdot\left[\varphi(\nabla\varphi\dot) \right] = \rho\bv\cdot\nabla\varphi+ \left[\frac{1}{2}(\nabla\varphi)^2  \bId -\nabla\varphi\circ\nabla\varphi \right]:\nabla\bv
}

The local Bondi balance is obtained by adding the local thermodynamic-Ohanian and Maxwell balances \re{tgravfbal_loc} and \re{Mgravfbal_loc} and dividing the result by two:
\eqn{Bgravfbal_loc}{
\frac{1}{2}\pt{(\rho \varphi)} +
\nabla\cdot\left[\rho\bv\varphi+ 
\frac{1}{2}\left(\varphi\pt{\nabla\varphi}-\nabla\varphi\pt{\varphi}\right)\right] = \rho\bv\cdot\nabla\varphi 
}
One can give a partially substantial Bondi balance, with partial time derivatives in the energy current density:
\eqn{Bgravfbal_subst}{
\frac{1}{2}\rho \dot\varphi +
\nabla\cdot\left[
\frac{1}{2}\rho\bv\varphi+ 
\frac{1}{2}\left(
\varphi\pt{\nabla\varphi}-\nabla\varphi\pt{\varphi}
\right) \right] = 
\rho\bv\cdot\nabla\varphi, 
}
The complete substantial form can be obtained by substituting the partial time derivatives in the current density with substantial ones. Then one can get an expression without an apparent physical insight: 
\eqn{Bgravfbal_sstubst}{
\frac{1}{2}\rho \dot\varphi +
\nabla\cdot\left[
\frac{1}{2}\rho\bv\varphi+ 
\frac{\varphi^2}{2}\left\{\left(
\frac{\nabla\varphi}{\varphi}\right)^\cdot-\bv\cdot\nabla\left(\frac{\nabla\varphi}{\varphi}\right)\right\} \right] = 
\rho\bv\cdot\nabla\varphi, 
}

Therefore, starting from substantial Ohanian energy balance with a source term with mechanical dissipation we have shown that the energy balances of Ohanian, Maxwell and Bondi are the same and meantime we have given them not only in a local form \re{enbal_1}, \re{enbal_0} and \re{enbal_1/2}, but also their substantial forms both with the power density of the gravitational field and with mechanical dissipation as source terms. The nine expressions are equivalent on-shell, and with the continuity equation as condition. 

Summarising the conclusions: 
\begin{itemize}
  \item Not only the energy densities of Maxwell, Ohanian and Bondi are equivalent, but their related energy balances as well. 
  \item The emerging different source terms, the power density of the gravitational force and the power density of gravitational pressure, are equivalent, because of the {classical holographic property} \re{chol_f}. However, every balance can be written with the preferable source term, see \re{tgravfbal_loc}, \re{Mgravfbal_loc} and \re{Bgravfbal_loc}. 
  \item The equivalence of the three balances is independent of the particular boundary conditions.
  More properly, the relevant boundary conditions are transformed according to the definition of the current density terms in the balances. 
  \item It is remarkable that the energy balance \re{tgravbal_subst}, the {classical holographic property} \re{chol_f} as well as the Poisson equation \re{Gauss-Poisson} are  general consequences of the Second Law of Thermodynamics. That is why this particular form was chosen as the starting point of our calculations. 

\end{itemize}

Several special conditions fundamentally influence the physical perception of the various energy balances above. Those are: the balance concepts (density, flux and source), the displacement current, the classical holographic property and in particular the Poisson equation, the fundamental field equation of nonrelativistic, Newtonian gravity. The substantial forms separate the comoving and rest frame forms: both in the time derivatives and in the current densities.  The identities, like the one for gravitational displacement vector, \re{-divptg=mass-flow-cont}, \eqn{fluxid}{\nabla\cdot \left(   \rho\bv + \pt{\nabla\varphi}\right) = 0}
Therefore, the gravitational displacement current becomes conductive mass current on-shell, that is assuming the Poisson equation as a background condition. A crucial question is the identification of the minimal set of conditions.

\section{Energy balance and field equations}


Balances are considered secondary in theories of physics, the primary concepts are the field equations. However, thermodynamics reverses this order: the energy balance is the primary, and the field equation is secondary, derived. Moreover, the energy density alone determines the other parts of the balance. The current density and the source term, the dissipated energy can be calculated in nonequilibrium thermodynamics. The simplest relevant method separates the currents from the source; therefore, it is called \emph{divergence separation}. It is simple, but very effective in weakly nonlocal theories  \cite{GroMaz62b,Mau06a}. Also, it is closely related to more rigorous methods of Second Law analysis, such as the Liu procedure \cite{Van23a,SzucsVan25m}. 

In the following, the method is demonstrated by the construction of a gravitational energy balance. The starting point is the Ohanian energy density \re{eps_0}. Then, calculating its substantial time derivative and using the balance of mass \re{massbalance} as a constraint one can separate a total divergence term and identify the balance form:
\eqn{calc_tgravbal_subst}{
\rho \left(\varphi+\frac{(\nabla\varphi)^2}{2\rho}\right)^{\bf \cdot} =
\rho\dot\varphi + \rho\frac{(\nabla\varphi)^2}{2}\left( -\frac{\dot\rho}{\rho^2}\right) +
\nabla\varphi(\nabla\varphi\dot) =\nnl {}
= \nabla\cdot\left[\dot\varphi \nabla\varphi \right] +
 \nabla\bv :\left[\frac{1}{2}(\nabla\varphi)^2 \bId 
 -\nabla\varphi\circ\nabla\varphi \right]
 + \dot\varphi(\rho-\Delta\varphi).
}
Here, the first term, the total divergence identifies an energy current density. The second term is the power due to the bracketed gravitational pressure, which is the source term of the energy balance. The last term, $\dot\varphi(\rho-\Delta\varphi)$, is the off-shell residual: it measures the deviation from the Poisson equation. Thermodynamic reasoning proves that in an ideal system, without dissipation, this term is zero and the Poisson equation follows --- the system is forced on-shell by the Second Law. The balance above is not the reason; it is a consequence.

One can transform the above equation into a local form, changing the substantial time derivative to partial one. They do not commute, therefore \re{stder} was used. Equivalently, one can perform the previous calculations, with the partial time derivative of the Ohanian energy density, also using the mass balance:
\eqn{calc_tgravfbal_loc}{
\partial_t \left(\rho\varphi
   + \frac{(\nabla\varphi)^2}{2}\right) =
\rho\pt{\varphi} + \varphi\pt{\rho} + \nabla\varphi\cdot\pt{\nabla\varphi} =\nonumber\\   
=\nabla\cdot\left[-\rho\bv\varphi + 
    \pt{\varphi }\nabla\varphi\right] +
\rho\bv\cdot\nabla\varphi + \pt{\varphi}(\rho-\Delta\varphi).
}
Therefore we have obtained the substantial balance of gravitational energy \re{tgravbal_subst} and also its local form \re{tgravfbal_loc} except that there is an additional source term: the Poisson equation multiplied by the time derivative according to the particular reference frame of the balance, the substantial or the local one. 
\\

{\color{black}
\subsection{The Green balance}

It was shown, that the three energy balances of gravitational energy are similar. However, the question is whether there is a difference in their physics?

Let us recall the differences between the energy densities. On any compact domain, the integral of the sum of the Maxwellian field energy density  
\eqn{edensMaxwell}{
\eps_1 = -\frac{1}{2}\bg^2,
}
and the Bondian matter energy density:
\eqn{edensBondi}{
\eps_{\frac{1}{2}} = + \frac{1}{2} \rho \varphi
}
is the same, because 
\eqn{int_identity}{
\int{(\rho\varphi+\bg^2)}\d V 
= \int \left(\varphi\Delta\varphi+\nabla\varphi\cdot\nabla\varphi\right) \d V
= \int \nabla\cdot \left( \varphi\nabla\varphi \right) \d V
= - \oint{\varphi\bg}\cdot \d{\bf A}.
}
where the Poisson equation \re{Gauss-Poisson} is used in the first equivalence, 
and the irrotationality condition \re{irrotcons} and the divergence theorem are used in the last one.
One may observe, that the supposed equivalence of different gravitation energies is based on the following combination $\nabla\cdot(\varphi\nabla\varphi)$, \re{int_identity}. Seemingly, the integration for the whole space and particular boundary conditions were the conditions. However, that argument is actually based on a strange conservation law.

First, we simply expand the leftmost $\nabla$ derivative: 
\eqn{expandnabla}{
\nabla\cdot(\varphi\nabla\varphi)
=
\varphi\Delta\varphi+\nabla\varphi\cdot\nabla\varphi
}

Then we take the time derivative, and exchange the time and space derivatives:
\eqn{ptswapnabla}{
\pt{}\nabla\cdot(\varphi\nabla\varphi) = \nabla\cdot\pt{}(\varphi\nabla\varphi)
}
This expression has a balance form, moreover, a balance form without a source term. Additionally, if the r.h.s. of \re{expandnabla} is inserted into the l.h.s. of \re{ptswapnabla}, 
and independently the time derivative on the r.h.s. is expanded and moved to the l.h.s., one obtains:
\eqn{ptLaplace_etc}{
\pt{(\varphi\Laplace\varphi+\nabla\varphi\cdot\nabla\varphi)} -
\nabla\cdot\left[(\pt{\varphi})\nabla\varphi + \varphi(\pt{\nabla\varphi})\right] =0.
}
In the following it will be called \emph{Green balance}, considering its origin.  

So far, identity \re{ptLaplace_etc} is off-shell: we used only general mathematical vector identities, in particular, neither the Poisson equation $\Delta \varphi = \rho$,
nor $\nabla \varphi = -\bg$.
Now we go on-shell, recalling NG and using both relations, and finally moving all r.h.s. terms to the l.h.s.:
\eqn{ptPoisson_etc}{
\partial_t\left(\rho\varphi+\bg^2\right) + 
\nabla\cdot\left[(\pt{\varphi})\bg + \varphi(\pt{\bg})\right] = 0
}
With these two independent physical inputs, the identity \re{ptLaplace_etc} takes the form of 
a sourceless local conservation law $\pt{\eps}+\nabla \cdot \mathbf{S}=0$, which is the time derivative of the Green identity $\nabla\cdot(\varphi\nabla\varphi) = \varphi\Delta\varphi + (\nabla\varphi)^2$, read as a balance.
In the energy current terms, we recognize (minus) the NG Maxwell displacement current $-\pt{\bg}$ times the scalar potential field $\varphi$,
and the gravitational variation $\pt{\varphi}$ times the gravitational field strength $\bg$.
} 

Let us step back to the off-shell form \re{ptLaplace_etc}. It is straightforward to transform it to a substantial form and obtain 
\eqn{sptLaplace_etc}{
\rho\left(\varphi\frac{\Laplace\varphi}{\rho}+
\frac{(\nabla\varphi)^2}{\rho}\right)^\cdot -
\nabla\cdot\left[(\varphi\nabla\varphi\dot) + \bv\cdot\left((\nabla\varphi)^2\bId - \nabla(\varphi \nabla\varphi)\right)\right] =0.
}
Here one can recognize the gravitational pressure in the current density. Let us separate the conditions in the transformations: \re{sptLaplace_etc}, is the substantial form of the {Green balance}. We have distinguished on-shell and off-shell local forms, \re{ptLaplace_etc} and \re{ptPoisson_etc}, respectively. The substantial form of the Green balance \re{ptLaplace_etc} requires the mass balance, but is off-shell because the Poisson equation was not a condition there.  

Now one can understand exactly the equivalence of the energy balances  \re{tgravfbal_subst}, \re{Mgravfbal_subst} and \re{Bgravfbal_subst}, as well as \re{tgravfbal_loc} \re{Mgravfbal_loc} and \re{Bgravfbal_loc}. They differ only in the {Green balance} above. For the substantial forms subtracting  \re{sptLaplace_etc} from \re{tgravfbal_subst} one obtains the Maxwell balance \re{Mgravfbal_subst} and subtracting half of \re{sptLaplace_etc} from \re{tgravfbal_subst} one obtains the substantial Bondi form \re{Bgravfbal_subst}.    Also, subtracting  \re{ptPoisson_etc} from  \re{tgravfbal_loc} yields the local Maxwell balance  \re{Mgravfbal_loc}, while subtracting half of  \re{ptPoisson_etc} from  \re{tgravfbal_loc} yields the local Bondi form  \re{Bgravfbal_loc}. 

{
Here, it is convenient to define the "Green energy density" $\eps_\nabla$:
\eqn{eps_nabla}{
\eps_\nabla
:=+\nabla\cdot\left[(\nabla\varphi)\varphi\right]
= -\nabla\cdot\left(\bg\varphi\right)
=-\bg\cdot(\nabla\varphi) -(\nabla\cdot\bg)\varphi
=+\bg^2 +\rho\varphi
}
and its corresponding "Green energy current density" $\bS_\nabla$:
\eqn{S_nabla}{
\bS_\nabla
:=-\pt{}\left[(\nabla\varphi)\varphi\right]
= +\pt{}\left(\bg\varphi\right)
= +\bg\left(\pt{\varphi}\right) +\left(\pt{\bg}\right)\varphi
= +\bS_\varphi -\bS_g 
}
where, in both \re{eps_nabla} and \re{S_nabla},
the first equality is the definition, the second equality uses the irrotationality condition \re{irrotcons}, the third equality is a mathematical manipulation and the fourth equality in \re{eps_nabla} uses the Poisson equation \re{Gauss-Poisson} (on shell).
Let us emphasize that, strictly speaking, 
$\eps_\nabla,\bS_\nabla$ are not gravitational energy and current, 
because there is no source term in their balance, which expresses conservation:
\eqn{eps_nabla_balance}{
\pt{\eps_\nabla} + \nabla\cdot\bS_\nabla =0,
}
 
More explicitly (and inserting an overall factor $1/2$ to ease comparison):
\eqn{enbal_nabla}{
\pt{}       \left(  \frac{1}{2}\bg^2 + \frac{1}{2}\rho\varphi\right)
+\nabla\cdot\left[+\frac{1}{2}\bg\left(\pt{\varphi}\right)  +\frac{1}{2}\left(\pt{\bg}\right)\varphi \right]
= 0
}
to be compared with the energy balances \re{enbal_1}, \re{enbal_0}, \re{enbal_1/2}.
Note that here $\nabla\cdot\left[\rho\bv\varphi\right]$ is absent on the l.h.s.,
and again $w=\rho\bv\cdot\nabla\varphi$ is absent on the r.h.s.
\\
It is now simple to prove that each corresponding pair $\eps_a,\bS_a$,
for any arbitrary value of the parameter $a$, 
can be expressed as a corresponding (i.e. with the same weights $1,-a$) 
linear combination of the reference energy density and current $\eps_0,\bS_0$ and 
 the {pseudo} energy density and current
$\eps_\nabla,\bS_\nabla$.
Indeed, comparing \re{eps_1}, \re{eps_0}, and \re{eps_nabla}
it is easy to see that $(\eps_1-\eps_0)=-\eps_\nabla$.
Rearranging \re{eps_a_lincomb} as $\eps_a=a \cdot(\eps_1-\eps_0)+\eps_0$, 
and substituting $(\eps_1-\eps_0)$ into it,
gives:
\eqn{eps_a_lincomb_nabla}{
\eps_a = \eps_0- a \cdot \eps_\nabla
}
Similarly, comparing \re{S_1}, \re{S_0}, and \re{S_nabla}
it is easy to see that $(\bS_1-\bS_0)=-\bS_\nabla$.
Rearranging \re{S_a_lincomb} as $\bS_a=a \cdot(\bS_1-\bS_0)+\bS_0$, and substituting $(\bS_1-\bS_0)$
gives:
\eqn{S_a_lincomb_nabla}{
\bS_a = \bS_0 - a \cdot \bS_\nabla
}
Note how these linear combinations \re{eps_a_lincomb_nabla} and \re{S_a_lincomb_nabla} are of a different kind 
with respect to the linear combinations \re{eps_a_lincomb} and \re{S_a_lincomb} previously used in the literature:
instead of using two terms modulated by two weights $a,(1-a)$ as those there, 
these here use only one, somewhat more "physical", reference block
(Ohanian's $\eps_0$, or $\bS_0$, with weight 1 i.e. fixed),
and another one, somewhat more "mathematical", "tunable" block 
(Green's $\eps_\nabla$, or $\bS_\nabla$, modulated by only one weight $a$).

As a consequence, we obtain a straightforward reinterpretation of the previously discussed parametrization of the energy balances, energy densities and energy current densities, shown in Sect.\re{sect:enbal}. The new parametrization is based on the Green balance and stresses the role played by the total divergence term $\nabla\cdot[(\nabla\varphi)\varphi]$: 
\eqn{balpar}{
Balance_a = Balance_0 - a \times Balance_{Green},
}
where the parameter "$a$" has the same values as above: 
$a=0$ for the Ohanian-No-Name, $a=1$ for the Maxwell-Livens and $a=1/2$ for the Bondi-Bondi balances, 
as well as for the energy density $\eps_a$ and energy current density $\bS_a$, respectively.

\section{Summary and discussion}

Balances are characteristic of classical field theories and continuum theories. An analysis of various energy densities and energy balances of Newtonian gravitation leads to the following conclusions.
\begin{enumerate}
\item The energy balances of Ohanian, Maxwell and Bondi are equivalent from a physical point of view: they differ only by the  {Green balance} \re{ptPoisson_etc}, the time derivative of the Green identity read as a conservation law. This equivalence holds locally, independently of any particular boundary conditions at spatial infinity. 
\item The bulk force power and the surface traction power source terms of the gravitational energy balance are equivalent because of the {classical holographic property} \re{chol_f}. 
\item The distinction of Eulerian and Lagrangian representations, the local and substantial forms is a key aspect for transforming the energy current density terms and for understanding the role of the gravitational displacement current density. One must distinguish between forms where the mass balance is the condition and when the Poisson equation is a that.
\end{enumerate}

Among the energy densities (and balances) the Ohanian form seems distinguished. It is independently supported by three different lines of reasoning. From the thermodynamic point of view, it emerges naturally from the Second Law analysis of a weakly nonlocal scalar field \cite{VanAbe22a,SzucsVan25m}. None of the other energy densities can be used in the thermodynamic theory (see Appendix \re{AppBalEnergy}).
From classical Lagrangian field theory, its opposite serves as the gravitational Lagrangian density from which the Euler--Lagrange equation yields the Poisson equation \cite{Fra15a, DewWea18a,BenEkl23a} (Appendix \re{AppLagrangian}). 

The coincidence between the thermodynamic and the variational (Lagrangian) selection of the Ohanian energy density is not accidental. Both rely on the off-shell structure: the matter-field coupling term $\rho\varphi$ in $\varepsilon_0$ is essential for generating the Poisson equation, whether through the Euler--Lagrange equation or through thermodynamic divergence separation. The  {Green balance} freedom, which mediates between the on-shell equivalent energy balances, cannot create or destroy this coupling term, since off-shell it is a pure mathematical identity involving $\varphi\Delta\varphi$, not $\rho\varphi$. Thus, the gauge-like freedom of the balances operates at a different (on-shell) level than the generating principle (off-shell), and the Second Law acts as a gauge-fixing that selects the unique generator.

The \emph{{Green balance}} \re{ptPoisson_etc}, the time derivative of the Green identity read as a sourceless conservation law. It represents a gauge-like freedom in the balance structure, distinct from the gauge ambiguity of the gravitational potential discussed by Dewar \& Weatherall \cite{DewWea18a}: their ambiguity resides in the field variable ($\varphi \to \varphi + f(t)$), whereas here the freedom acts at the level of the energy balance itself. Adding $a$ times the {Green balance} to the Ohanian balance redistributes energy between density and current without affecting the source term. Off-shell, the Green identity is purely mathematical; only when the Poisson equation holds can its terms be identified as the difference of physical energy densities, 
$\varepsilon_0 - \varepsilon_1 = \rho\varphi + (\nabla\varphi)^2 
= \nabla\cdot(\varphi\nabla\varphi)$, 
rendering them equivalent upon integration. The thermodynamic approach 
does not presuppose this on-shell condition but derives it: the Second 
Law, applied to the Ohanian energy density via divergence separation, 
forces $\Delta\varphi=\rho$ as the dissipation-free limit. In this sense, 
the Second Law acts as a physical gauge-fixing principle that selects the 
Ohanian balance from the family of equivalent alternatives.

A more complete four-dimensional energy-momentum balance in Galilean relativistic continuum theory may provide further insight, in particular regarding this gauge-like aspect as a hidden feature of balance-based continuum theories. However, the covariant energy balance in the Galilean relativistic framework is surprisingly complex and conceptually different from the traditional scalar formula: the kinetic energy becomes a relative transformation property due to the four-tensor transformation properties of the energy-momentum tensor \cite{Mat20b}. This direction remains open for future investigation.

%

\begin{appendices}

\section{The general balance of energy}
\label{AppBalEnergy}

In the field theoretic framework of nonequilibrium thermodynamics there is a methodology for constructing evolution equations. It was developed for modeling properties of dissipative and perfect materials with the help of internal variables \cite{Ver97b}. The evolution equations cannot violate the Second Law of Thermodynamics and must be compatible with any given constraints, including the fundamental balances. The simplest version of the methodology was introduced by de Groot and Mazur \cite{GroMaz62b}. Here, the method is applied directly to construct the energy balance, without the thermodynamic background, as it is extracted and simplified from the calculations of \cite{SzucsVan25m}.

The key of connecting the mechanical and thermodynamic concepts is internal energy. In general entropy is the function of the total energy and the various other field. Internal energy is defined by subtracting the identified energies from the total energy. Internal energy that is distributed evenly in thermodynamic equilibrium considering all other constraints. If the momentum balance and gravitational field equation are the constraints, then one subtracts the kinetic energy and also the energy of gravity from the total energy and obtains the internal energy. In classical irreversible thermodynamics the difference of the total, conserved energy and the kinetic energy gives the classical form of the internal energy. Then the entropy depends on this classical internal energy $e$, and also on the other field variables in a special way:
$\rho_s(\rho_{tote}, \rho, \varphi, \nabla\varphi) = \hat \rho_s(\rho_{tote} - \varepsilon(\rho, \varphi, \nabla\varphi), \rho)$ where $ \varepsilon$ denotes the energy density of the gravitation. Then it is clear, how the entropy balance is connected to the energy balance \cite{CimEta14a,SzucsVan25m}. 

Then let us consider the specific (\textit{i.e.}, per unit mass) energy $e = \frac{\varepsilon}{\rho}$ as a general function of the gravitational potential, its gradient and the mass density
\begin{equation}
\label{e}
e=e(\rho, \varphi, \nabla \varphi):=\frac{\eps(\rho, \varphi, \nabla \varphi)}{\rho}
\end{equation}
Now we restrict ourselves calculate its balance, separating the surface and production terms, where the compatibility with the balance of mass is required. First, the mass balance is considered in the following substantial form:
\begin{gather}
    \dot{\rho} + \rho \nabla \cdot \mathbf{v} = 0.
\end{gather}
We expand the differential of $e$ as 
\begin{gather}
    \mathrm{d} e = \partial_\rho e \mathrm{d}\rho + \partial_\varphi e \mathrm{d}\varphi + \partial_{\nabla \varphi }e \mathrm{d}(\nabla \varphi),
\end{gather}  
where the coefficients of the differentials are the following partial derivatives:
\begin{gather}
     \partial_\rho e :=\left. \frac{\partial e}{\partial \rho}\right|_{ \varphi, \nabla \varphi}, \nonumber \\
    \partial_\varphi e :=\left. \frac{\partial e}{\partial \varphi}  \right|_{\rho,  \nabla \varphi} , \nonumber \\
     \partial_{\nabla \varphi} e:=  \left. \frac{\partial e}{\partial (\nabla \varphi)} \right|_{\rho, \varphi}.\nonumber
\end{gather}

Calculating the substantial time derivative of the specific energy $\dot{\varepsilon}$, the following familiar balance form emerges:
\begin{equation}
\rho \dot{e} + \nabla  \cdot\mathbf{j}_e = - \mathbf{P}:\nabla\mathbf{v}.    
\end{equation}
Thus, we rewrite $\rho \dot{e}$ using the above equations with this derivation:
\begin{gather}
    \rho \dot{e} = \rho \partial_\rho e\dot{\rho} + \rho \partial_\varphi e \dot{\varphi}+\rho \partial_{\nabla \varphi} e (\nabla \varphi)^{\cdot}= \\ \nonumber
    = -\rho^2 \partial_\rho e \nabla \cdot \mathbf{v} + \rho \partial_\varphi e \dot{\varphi}+\rho \partial_{\nabla \varphi}  e (\nabla \dot{\varphi} - (\nabla \varphi)\cdot(\nabla \mathbf{v}))= \\ \nonumber
     = -\rho^2 \partial_\rho e \mathbf{I}: (\nabla \mathbf{v}) + \rho \partial_\varphi e \dot{\varphi}+\rho \partial_{\nabla \varphi}  e \cdot (\nabla \dot{\varphi}) - \rho (\partial_{\nabla \varphi}  e)  (\nabla \varphi):(\nabla \mathbf{v}). 
     \end{gather}
     Therefore, with some additional rearrangements, the following general form is obtained
\begin{gather}
    \rho \dot{e} + \nabla \cdot \left[ - \rho \dot{\varphi} \partial_{\nabla \varphi} e \right] = \nonumber\\ \qquad 
    \dot{\varphi}(\rho \partial_\varphi e -\nabla \cdot (\rho  \partial_{\nabla\varphi}e)) - \left[\rho^2 \partial_\rho e \mathbf{I} + \rho (\partial_{\nabla \varphi} e)(\nabla \varphi) \right]:(\nabla \mathbf{v}). \label{energy_bal_sep}
\end{gather}
Then, the last term is the mechanical dissipation. In the penultimate term, one may recognize a functional derivative as the coefficient of the substantial time derivative of the gravitational potential, and the current density of the balance is also identified.  

Alternatively, instead of using the substantial time derivative, one can perform similar calculations with partial time derivatives and obtain the local form of the energy balance (now using the local form of mass balance, eq. \re{mass-flow-cont}, as well):
\begin{gather}
    \partial_t (\rho e) = \partial_\rho(\rho e) \partial_t \rho +  \partial_\varphi (\rho e) \partial_t{\varphi} +
    \partial_{\nabla \varphi}(\rho e) \partial_t(\nabla \varphi) = \nonumber\\
    =-\partial_\rho(\rho e) \nabla\cdot(\rho \mathbf{v})+  
    \partial_\varphi (\rho e) \partial_t{\varphi} +
    \nabla\cdot( \partial_{\nabla \varphi}(\rho e) \partial_t\varphi) -
     \nabla\cdot(\partial_{\nabla \varphi}(\rho e)) \partial_t\varphi =\nonumber\\
    \nabla\cdot( \partial_{\nabla \varphi}(\rho e) \partial_t\varphi- \rho \mathbf{v}\partial_\rho(\rho e))+
     ( \partial_\varphi (\rho e)-
        \nabla\cdot(\partial_{\nabla \varphi}(\rho e)) \partial_t\varphi+
     \rho \mathbf{v}\cdot\nabla\cdot\partial_\rho(\rho e).
\nonumber\end{gather}
 Regrouping the terms one obtains
 \begin{gather}
    \partial_t(\rho e) +\nabla \cdot \left[\rho \partial_{\rho}(\rho e) \mathbf{v}- \rho \partial_t \varphi \partial_{\nabla \varphi} e \right]=  \partial_t{\varphi} \left(\rho \partial_\varphi e  - \nabla \cdot (\rho  \partial_{\nabla \varphi} e  ) \right) + 
    \rho \mathbf{v} \cdot \nabla (\partial_{\rho}(\rho e)),
\label{locebal}\end{gather}
as the general form of eq. \re{tgravfbal_loc} with extra terms. Here the coefficient of the partial time derivative of $\varphi$ is the functional derivative of the energy density $\rho e$ and the last term is the power of the force density $\rho \nabla (\partial_{\rho}(\rho e))$. It is straightforward to check that \re{locebal} is identical to \re{energy_bal_sep}, because 
\begin{gather}
\nabla\cdot\left[\rho^2 \partial_\rho e \mathbf{I} + \rho (\partial_{\nabla \varphi} e)(\nabla \varphi) \right] = \rho  \nabla (\partial_{\rho}(\rho e))
\end{gather}
The divergence of the gravitational pressure equals the power of the force density: the {\color{black}classical holographic property} in general form. 

The most important, and somewhat hidden property of the equations above is that the last but one term in the balances \re{energy_bal_sep} and \re{locebal}, represents the off-shell dissipation. The on-shell form, with the field equation defined by the zero functional derivative $\delta_\varphi e =0$, represents zero dissipation. The thermodynamic argument shows that it is a dissipative term. Therefore, if gravity is nondissipative, the term must be zero, regardless of whether the gravitational potential is increasing or decreasing. We refer to \cite{SzucsVan25m} regarding the detailed justification.

Now we specialize the specific energy \re{e} to the cases coreesponding to the gravitational energy densities
discussed in the main text: 
Maxwell's $\eps_1$ eq.\re{eps_1}, Ohanian's $\eps_0$ eq.\re{eps_0}, Bondi's $\eps_{\frac{1}{2}}$ eq.\re{eps_1/2}.
In the Ohanian case, with
\begin{equation}
e_0(\rho, \varphi, \nabla \varphi) 
:= \frac{\eps_0(\rho,\varphi,\nabla\varphi)}{\rho}
:= \frac{\rho \varphi + \frac12 (\nabla \varphi)^2}{\rho},
\end{equation}
substituted in eq.\re{energy_bal_sep} we obtain the following form:
\begin{equation}
    \rho \dot{e}_0 + \nabla \cdot \left[ -  \dot{\varphi} \nabla \varphi \right] =  \dot{\varphi}( \rho  - \Delta \varphi) - \left[(\nabla \varphi) (\nabla \varphi)- \frac12 (\nabla \varphi)^2  \mathbf{I}   \right]:(\nabla \mathbf{v}). \label{e_bal_sep_ohanian}
\end{equation}

Alternatively, with the local form from eq. \re{locebal} one gets:
\begin{equation} \label{e_bal_sep_loc_ohanian}
\partial_t \left( \rho e_0\right)     +  \nabla \cdot \left[ \rho \varphi \mathbf{v} - \partial_t \varphi \nabla \varphi \right] =  \partial_t \varphi \left(\rho - \Delta \varphi \right) + \rho \mathbf{v} \cdot \nabla \varphi.
\end{equation}

So we have obtained the off-shell form of \re{tgravbal_subst} and \re{tgravfbal_loc} as a consequence of the general off-shell balances.  The Ohanian energy density generates the complete energy balance.

Instead, using the specific Maxwell's energy, with
\begin{equation}
e_1(\rho, \varphi, \nabla \varphi)
:= \frac{\eps_1(\nabla\varphi)}{\rho}
:= \frac{- \frac12 (\nabla \varphi)^2}{\rho},
\end{equation}
substituted in eq. \re{energy_bal_sep} we obtain the following form:
\begin{equation}
    \rho \dot{e}_1 + \nabla \cdot \left[   \dot{\varphi} \nabla \varphi \right] =  \dot{\varphi} \Delta \varphi  - \left[ \frac12 (\nabla \varphi)^2  \mathbf{I}   -(\nabla \varphi) (\nabla \varphi)\right]:(\nabla \mathbf{v}), \label{e_bal_sep_maxwellian}
\end{equation}
and from eq. \re{locebal}:
\eqn{distmaxwen}{
     \partial_t \left(\rho{e}_1 \right)  + \nabla \cdot \left[ \partial_t\varphi \nabla \varphi\right] =  \partial_t \varphi \Delta \varphi.
}
One can see that the energy balances of the gravitational energy with the Maxwell energy density \re{MgravPbal_subst} and \re{Mgravfbal_loc} look like different.

The Bondi case, with
\begin{equation}
e_\frac{1}{2}(\rho, \varphi, \nabla \varphi) 
:= \frac{\eps_\frac{1}{2}(\rho,\varphi)}{\rho}
:= \frac{+\frac12 \rho \varphi}{\rho}
=\frac12 \varphi
\end{equation}
substituted in eq. \re{energy_bal_sep} we obtain the following trivial identity:
\begin{equation}
    \rho \dot{e}_\frac{1}{2}=  \frac12 \rho \dot{\varphi}, \label{e_bal_sep_bondi}
\end{equation}
and from eq. \re{locebal}:
\begin{equation}
     \partial_t \left(  \rho \varphi \right)  + \nabla \cdot \left[ \rho \mathbf{v}  \varphi \right] 
     =    \rho \partial_t \varphi  +  \rho \mathbf{v} \cdot \nabla \varphi .
\end{equation}
It is even more strange than the Maxwell balance. None of them 
result in a usual form of the energy balance, while the Ohanian energy density leads to a complete balance without a preliminary knowledge of the field equation. Moreovere, it is a generator of the field equation and the gravitational pressure in a thermodynamic framework \cite{VanAbe22a,SzucsVan25m}.

Thus, one can see that the simplified thermodynamic procedure distinguishes between the various energy densities:
the Second Law acts as a physical gauge-fixing principle. 

\section{About the generalisation of the {Green balance}}
\label{AppBalq}

One can start with the following identity 
\eqn{tbal}{
\pt{}\nabla\cdot \bf{q}  = \nabla\cdot\pt{ \bf{q} }
 }
 where $\bf{q}$ is a contravariant vector field.  
The upper convected time derivative of $\bf q$ is 
\eqn{uppconv}{
 \bf{q}^\nabla = \pt{\bf q} + \bv\cdot\nabla \bf{q} - \bf{q}\cdot\nabla\bv =
  \dot{\bf q}  - \bf{q}\cdot\nabla\bv
}
The Truesdellian time derivative of $\bf q$ is  \cite{BamMor80a}
\eqn{uppconv_Truesdal}{
 \bq^\diamond:= \bq^\nabla+ \bq\nabla\cdot\bv
}

Then it is easy to see, that \re{tbal} can be rewritten as
\eqn{substbal}{
(\nabla\cdot\bq\dot) -\nabla\cdot\bq^\nabla - \bq\cdot\nabla(\nabla\cdot\bv) =
\rho\left(\frac{\nabla\cdot\bq}{\rho}\right)^\cdot -\nabla\cdot\bq^\diamond
=0.
}
In index notation (with  upper an  lower  indexes denoting the contravariant and covariant components respectively):
\eqn{substbal_idx}{
(\partial_i q^i)^\cdot - \partial_i \left(\dot q^i - q^k \partial_k v^i\right) - q^i \partial_i(\partial_k v^k) =
\rho\left(\frac{\partial_i q^i}{\rho}\right)^\cdot - \partial_i\left(\dot q^i - q^k \partial_k v^i + q^i \partial_k v^k\right)
=0.
}
Here the continuity equation was used.

\section{Ohanian's energy density vs Lagrangian field theory}
\label{AppLagrangian}

Remarkably, \emph{Ohanian's energy density \re{eps_0} equals (the opposite of)}
$\mathcal{L}_g$, the well-known "gravitational" Lagrangian density
(see e.g. \cite{DewWea18a} eq.(1), \cite{Ekl22a} Sect.1.4.0, 
or \cite{Fra15a} Sect.II.B):
\eqn{L_g}{
\mathcal{L}_g(\varphi,\nabla\varphi|\rho)
:= -\left(\varphi\rho + \frac{1}{2}(\nabla\varphi)^2 \right)
= - \eps_0(\rho,\varphi,\nabla\varphi)}
Substituting $\mathcal{L}_g$ in the Euler-Lagrange equation:
\eqn{Euler-Lagrange}{
\frac{\partial \mathcal{L}_g}{\partial \varphi}
= \nabla\cdot \left( \frac{\partial \mathcal{L}_g}{\partial (\nabla\varphi)} \right)
}
gives the two field equations of Newtonian Gravitation, namely
the Gauss-Poisson eq.\re{Gauss-Poisson} and the irrotationality condition condition \re{irrotcons}.

In the usual physical interpretation, 
the pure gravitational term $\frac{1}{2}(\nabla\varphi)^2$ describes the gravitational field "on its own",
while the coupling" term $\rho\varphi$ describes the "inter-action" of matter $\rho$ and gravity $\varphi$.
However, note that in $\mathcal{L}_g$ in \re{L_g}
one is assuming a mass density distribution $\rho(\bx,t)$=fixed, given a priori,
regarded as a "parametric variable (function)", but not as a dynamically variable field itself.
Indeed, the partial derivatives of $\mathcal{L}_g$ in \re{Euler-Lagrange} are only w.r.t. 
the fields $\varphi$ and $\nabla\varphi$, but not w.r.t. $\rho$.
Therefore, in this context, 
modifying $\rho$ "by hand" obviously modifies the resulting $\varphi$,
instead varying $\varphi$ to satisfy the Euler-Lagrange equation does not modify $\rho$ already fixed a priori.
So to say: in this context
the "inter-action" is only "action" (the generation of gravitation from matter, i.e. the Gauss-Poisson field equation)
but there is no corresponding "back-reaction" (the equations of motions of matter under the force of gravity).
Paraphrasing a famous statement usually attributed to John Archibald Wheeler:
"matter tells gravity how to spread in space, but gravity does not (yet) tell matter how to move".

The l.h.s. partial derivative $\frac{\partial \mathcal{L}_g}{\partial \varphi}$ 
is sometimes called the canonical (Lagrangian, generalized) force corresponding 
to the canonical (Lagrangian, generalized) coordinate $\varphi$:
\eqn{LagrF}{
\mathbf{F}_\varphi:=\frac{\partial \mathcal{L}_g}{\partial\varphi}= -\rho
}
while the r.h.s. partial derivative $\frac{\partial \mathcal{L}_g}{\partial (\nabla\varphi)}$
is called the canonical (Lagrangian, generalized) momentum conjugated to $\varphi$:
\eqn{LagrP}{
\mathbf{P}_\varphi:=\frac{\partial \mathcal{L}_g}{\partial (\nabla\varphi)}=-\nabla\varphi=+\bg
}
Substituting \re{LagrF} and \re{LagrP} into \re{Euler-Lagrange}
gives the Gauss-Poisson field equation \re{Gauss-Poisson}:
\eqn{Gauss-Poisson-Eluer-Lagrange}{
-\rho
=\nabla\cdot(-\nabla\varphi)
=+\nabla\cdot\bg
}
The other field equation \re{irrotcons} of NG, the irrotationality condition $\nabla\times\bg=0$,
is already implied by:
(i) the physical assumption of $\varphi$ 
as an appropriate field "coordinate" in \re{Euler-Lagrange},
(ii) the mathematical definition \re{g=-gradphi} of $\bg=-\nabla\varphi$ 
as an "auxiliary" vector field, 
derived from the more fundamental scalar field $\varphi$, ass assumed in (i), and
(iii) the mathematical vector identity $\nabla\times\nabla\equiv0$.

Note that, if one chooses an arbitrary energy density $\eps_a$ 
(among those compatible with the energy balances) in place of $\eps_0$,
and then repeats the above steps from \re{L_g} to \re{Gauss-Poisson-Eluer-Lagrange}, one gets instead:
\eqn{Gauss-Poisson-Eluer-Lagrange_a}{
-(1-a)\rho
=+(1-2a)\nabla\cdot\bg
}
Comparing this "Poisson-like" equation \re{Gauss-Poisson-Eluer-Lagrange_a}
with the true Poisson equation \re{Gauss-Poisson-Eluer-Lagrange},
and assuming in general $\rho\not=0$ and $\nabla\cdot\bg\not=0$,
gives $1-a=1-2a$, with unique remaining solution $a=0$.
In particular, Maxwell's $\eps_1$ gives $0=\nabla\cdot\bg$ 
(which might apply to the gravitational field $\bg$, but only in empty space $\rho=0$)
and Bondi's $\eps_\frac{1}{2}$ gives $-\frac{1}{2}\rho=0$ 
(which also applies only to empty space, but furthermore it is totally silent about the gravitational field $\bg$).
So, in this context Ohanian's energy density $\eps_0$ is definitely distinguished, 
because it emerges immediately (with no other assumptions) from the initial expression for $\mathcal{L}_g=-\eps_0$,
while its two direct rivals $\eps_1$ and $\eps_\frac{1}{2}$ perform rather poorly.

\end{appendices}


\section{Acknowledgement}   
The authors acknowledge networking support by the grant NKFIH NKKP-Advanced 150038 and STSM Grant from FuSe COST Action [CA24101], funded by COST (European Cooperation in Science and Technology). 

The research reported in this paper is part of project no. TKP-6-6/PALY-2021, implemented with the support provided by the Ministry of Culture and Innovation of Hungary from the National Research, Development and Innovation Fund, financed under the TKP2021-NVA funding scheme.

We thank the Galileo Galilei Institute for Theoretical Physics for hospitality and the INFN for partial support during the completion of this work.

\bibliographystyle{unsrt}


\end{document}